# On the Computation of the Higher-Order Statistics of the Channel Capacity over Generalized Fading Channels

Ferkan Yilmaz, *Member, IEEE* and Mohamed-Slim Alouini, *Fellow, IEEE*

*Abstract*—The higher-order statistics (HOS) of the channel capacity $\mu_n = \mathbb{E}\left[\log^n\left(1 + \gamma_{end}\right)\right]$, where $n \in \mathbb{N}$ denotes the order of the statistics, has received relatively little attention in the literature, due in part to the intractability of its analysis. In this letter, we propose a novel and unified analysis, which is based on the moment generating function (MGF) technique, to exactly compute the HOS of the channel capacity. More precisely, our mathematical formalism can be readily applied to maximal-ratio-combining (MRC) receivers operating in generalized fading environments. The mathematical formalism is illustrated by some numerical examples focusing on the correlated generalized fading environments.

## I. Introduction

ERGODIC capacity, i.e., $\mu = \mathbb{E}\left[\log\left(1 + \gamma_{end}\right)\right]$, where $\mathbb{E}\left[\cdot\right]$ denotes the expectation operator and $\log(\cdot)$ denotes the natural logarithm, has been extensively investigated for diversity receivers in the literature (see for example [1]–[5] and the references therein). It is worth noticing that the *first-order statistics* of channel capacity is well-known as the ergodic capacity. However, the *higher-order statistics* (HOS) of the channel capacity $\mu_n = \mathbb{E}\left[\log^n\left(1 + \gamma_{end}\right)\right]$, where $n \in \mathbb{N}$ denotes the order of the statistics, are also useful in order to control the maximum dispersion in the channel capacity and achieve a successful carrier aggregation [6]. Some papers addressed the HOS of the channel capacity for different types of fading channels [4], [7, and references therein]. In particular, while the references in [7] have been involved with multiple input multiple output (MIMO) transmission over Rayleigh or Riciean fading channels, Laourine *et al.* considered in [7] the HOS of the channel capacity only for single-link log-normal fading channels. However due in part to the difficulty of finding a tractable analytical solution for diversity receivers, there has been indeed a lack of unified analysis in the literature on the HOS of the channel capacity.

Consider the instantaneous SNR $\gamma_{end}$ at the output of the $L$-branch maximal ratio combining (MRC), specifically given by $\gamma_{end} = \sum_{\ell=1}^{L} \gamma_{\ell}$ where for $\ell \in \{1, 2, \ldots, L\}$, $\gamma_{\ell}$ represents the $\ell$th diversity branch's instantaneous SNR. Without loss of generality, the HOS of the channel capacity for the $L$-branch MRC receiver is given by[1]

$$\mu_n = \underbrace{\int_0^\infty \int_0^\infty \cdots \int_0^\infty}_{L\text{-fold}} \log^n\left(1 + \sum_{\ell=1}^L \gamma_\ell\right) \times \\ p_{\gamma_{end}}\left(\gamma_1, \gamma_2, \ldots, \gamma_L\right) d\gamma_1 d\gamma_2 \ldots d\gamma_L, \quad (1)$$

where $p_{\gamma_{end}}\left(\gamma_1, \gamma_2, \ldots, \gamma_L\right)$ denotes the joint probability density function (PDF) of the instantaneous SNRs $\gamma_1, \gamma_2, \ldots, \gamma_L$. Note that this $L$-fold integration is tedious and computationally inefficient even if $\gamma_1, \gamma_2, \ldots, \gamma_L$ are assumed to be mutually independent. Recently, Di Renzo *et al.* derived in [4, Eq. (22)] an analytical solution to evaluate the HOS of the channel capacity. However, as mentioned in [4, Section III-E], this analytical solution requires both higher-order differentiation and a limit operation.

In this paper, we propose an MGF-based approach for the exact and unified analysis of the HOS of the channel capacity over correlated/uncorrelated generalized fading channels and for an MRC receiver with an arbitrary number of diversity branches. More specifically, in contrast to [4, Eq. (22)], our MGF-based approach does not require higher-order differentiation and limit operation in the numerical evaluation. It is further clear that, with the aid of Fox's H transforms [10], our MGF-based approach readily results in a closed-form for single-link fading channels. On the subject of the first-order statistics (i.e., ergodic capacity), our MGF-based approach also provides a new MGF-based approach for the ergodic capacity, which is alternative to the other MGF-based approaches proposed in [3], [4], and which may be useful for the researchers working in this field. Finally, numerical and simulation results, performed to verify the correctness of the proposed approach, are shown to be in perfect agreement.

## II. Higher-Order Statistics of Channel Capacity

In this section, a novel MGF-based framework for the exact analysis of the HOS for the channel capacity is introduced

---

Manuscript received April X, 2012. The associate editor coordinating the review of this letter and approving it for publication was X. XXXXXX.

Ferkan Yilmaz and Mohamed-Slim Alouini are with King Abdullah University of Science and Technology (KAUST), Al-Khawarizmi Applied Math. Building, Thuwal 23955-6900, Makkah Province, Kingdom of Saudi Arabia (e-mail: {ferkan.yilmaz, slim.alouini}@kaust.edu.sa).

This work was supported by King Abdullah University of Science and Technology (KAUST).

Digital Object Identifier 10.1109/WCL.031512.XXXXXX

[1]Note that if the PDF of $\gamma_{end}$ could be expanded in a canonical exponential form, then (1) is very easy and straightforward to compute, evidently leading to closed-form results. For instance, the HOS of the channel capacity have been successfully carried out in [8] for the diversity receivers over Nakagami-$m$ fading channels since the PDF of the Nakagami-$m$ distribution can be expanded in canonical form [9]. However for the other fading distributions commonly used in the literature (e.g, extended generalized-K and its special cases), the PDF of $\gamma_{end}$ is generally not available in a simple and canonical form.

for an MRC receiver in correlated/uncorrelated generalized fading environments.

**Theorem 1.** *The HOS of the channel capacity $\mu_n = \mathbb{E}\left[\log^n\left(1+\gamma_{end}\right)\right]$ for the L-branch MRC receiver over correlated fading channels is given by*

$$\mu_n = \int_0^\infty \mathcal{Z}_n(s)\left\{\mathcal{M}_{\gamma_{end}}(s) - \frac{\partial}{\partial s}\mathcal{M}_{\gamma_{end}}(s)\right\}ds, \quad (2)$$

*where $\mathcal{M}_{\gamma_{end}}(s) = \mathbb{E}\left[\exp\left(-s\sum_{\ell=1}^L \gamma_\ell\right)\right]$ denotes the joint MGF for the correlated instantaneous SNRs $\gamma_1, \gamma_2, \ldots, \gamma_L$. Furthermore, $\mathcal{Z}_n(s)$ is an auxiliary function defined as*

$$\mathcal{Z}_n(s) = n!\exp(-s)\,\mathrm{G}_{n+2,n+1}^{0,n+1}\left[\frac{1}{s}\left|\begin{array}{c}1,1,\ldots,1\\0,0,\ldots,0\end{array}\right.\right], \quad (3)$$

*where $\mathrm{G}_{p,q}^{m,n}[\cdot]$ denotes Meijer's G function [11, Eq. (9.301)].*

*Proof:* The proof has two steps, one of which is to establish (2) for the HOS of the channel capacity in correlated fading environments. The other one is to obtain the auxiliary function $\mathcal{Z}_n(s)$ in a closed form without involving any differentiation as compared to [4, Eq. (22)].

*The First Step.* Note that using the well-known relation between the differentiation and the power of logarithm, i.e., $\frac{\partial^n}{\partial a^n}(1+\gamma_{end})^{-a} = (-1)^n(1+\gamma_{end})^{-a}\log^n(1+\gamma_{end})$ where $n \in \mathbb{N}$ and $a \in \mathbb{R}^+$, the HOS of the channel capacity can be readily written as

$$\mu_n = \int_0^\infty \left\{(-1)^n\frac{\partial^n}{\partial a^n}(1+\gamma)^{-a}\Big|_{a=0}\right\}p_{\gamma_{end}}(\gamma)d\gamma. \quad (4)$$

Substituting the following equality [11, Eq. (1.512/4)], i.e.,

$$(1+\gamma)^{-a-1} = \frac{1}{\Gamma(a+1)}\int_0^\infty \exp(-s\gamma)s^a\exp(-s)\,ds \quad (5)$$

into (4) and performing some simple algebraic manipulations while exploiting the definition of joint MGF, i.e., $\mathcal{M}_{\gamma_{end}}(s) = \int_0^\infty \exp(-s\gamma)p_{\gamma_{end}}(\gamma)d\gamma$, we can readily obtain the HOS of the channel capacity $\mu_n = \mathbb{E}\left[\log^n(1+\gamma_{end})\right]$ as shown in (2), which proves the first step of Theorem 1, and the resulting auxiliary function $\mathcal{Z}_n(s)$ is given by

$$\mathcal{Z}_n(s) = (-1)^n \exp(-s)\frac{\partial^n}{\partial a^n}\frac{s^a}{\Gamma(a+1)}\bigg|_{a=0}, \quad (6)$$

which will be obtained in a closed form in the following step.

*The Second Step.* In order to simplify the auxiliary function $\mathcal{Z}_n(s)$, we use the well-known Cauchy's integral formula [12]. Note that, with the aid of Cauchy's integral formula, (6) can be expressed in terms of Mellin-Barnes Integral as

$$\mathcal{Z}_n(s) = n!\,e^{-s}\left\{\frac{1}{2\pi \mathrm{i}}\oint_{\mathcal{C}}\frac{\Gamma^n(-p)\,s^p}{\Gamma^{n+1}(1-p)\Gamma(1+p)}dp\right\}, \quad (7)$$

where $\mathrm{i} = \sqrt{-1}$ denotes the imaginary number and the contour integration $\mathcal{C}$ is chosen counter-clockwise in order to ensure the convergence. With the aid of the Mellin-Barnes integral representation of Meijer's G function [11, Eq. (9.301)], the Mellin-Barnes integral in (7) can be expressed in terms of Meijer's G function, that is

$$\frac{1}{2\pi \mathrm{i}}\oint_{\mathcal{C}}\frac{\Gamma^n(-p)\,s^p\,dp}{\Gamma^{n+1}(1-p)\Gamma(1+p)} = \mathrm{G}_{n+2,n+1}^{0,n+1}\left[\frac{1}{s}\left|\begin{array}{c}1,1,\ldots,1\\0,0,\ldots,0\end{array}\right.\right]. \quad (8)$$

Finally, substituting (8) into (7) yields the desired result given in (3), which proves the second step of Theorem 1. ∎

Note that Theorem 1 is applicable for any situation if the MGF of the instantaneous SNR $\gamma_{end}$ could be obtained. In addition, in the case of there does not exist any correlation between all diversity branches, the HOS of the channel capacity $\mu_n$ can be readily obtained using the following corollary.

**Corollary 1.** *When the L-branch MRC diversity receiver's diversity branches are not correlated, the HOS of the channel capacity $\mu_n$ is given by* [2]

$$\mu_n = \int_0^\infty \mathcal{Z}_n(s)\Bigg\{\prod_{\ell=1}^L \mathcal{M}_{\gamma_\ell}(s) - \\ \sum_{\ell=1}^L\left[\frac{\partial}{\partial s}\mathcal{M}_{\gamma_\ell}(s)\right]\prod_{\substack{k=1\\k\neq\ell}}^L \mathcal{M}_{\gamma_k}(s)\Bigg\}ds, \quad (9)$$

*where for all $\ell \in \{1,2,\ldots,L\}$, $\mathcal{M}_{\gamma_\ell}(s) = \mathbb{E}\left[\exp(-s\gamma_\ell)\right]$, $\Re\{s\} \in \mathbb{R}^+$ denotes the MGF of the $\ell$th diversity branch.*

*Proof:* When there is no correlation between all instantaneous SNRs $\gamma_1, \gamma_2, \ldots, \gamma_L$, one can readily write the joint MGF $\mathcal{M}_{\gamma_{end}}(s) = \prod_{\ell=1}^L \mathcal{M}_{\gamma_\ell}(s)$. The proof is thence obvious using the derivative of this joint MGF $\mathcal{M}_{\gamma_{end}}(s)$. ∎

In the following subsections, we consider some special cases in order to demonstrate the analytical simplicity and accuracy.

### A. Some Special Cases

The first-order statistics of the channel capacity, which is well-known as the ergodic capacity, have been greatly and extensively studied in the literature [1]–[4]. Thence, substituting $n = 1$ into (3) and exploiting the equality $\mathrm{G}_{3,2}^{0,2}\left[x\left|\begin{array}{c}1,1,1\\0,0\end{array}\right.\right] = -\mathrm{E}-\log(x)$ for all $x \in \mathbb{R}$, which we obtained with the aid of [13, Eqs. (07.34.03.0191.01) and (07.25.03.0005.01)], the auxiliary function $\mathcal{Z}_1(s)$ simply reduces to

$$\mathcal{Z}_1(s) = -\exp(-s)\left(\log(s) + \mathrm{E}\right), \quad (10)$$

where the constant $\mathrm{E} = 0.5772156649015328606\ldots$ is Euler-Mascheroni constant [11]. With this result, the ergodic capacity can be readily given by

$$\mu_1 = -\int_0^\infty \exp(-s)\left(\log(s) + \mathrm{E}\right) \times \\ \left\{\mathcal{M}_{\gamma_{end}}(s) - \frac{\partial}{\partial s}\mathcal{M}_{\gamma_{end}}(s)\right\}ds, \quad (11)$$

Given the advantages of the ergodic capacity analysis in [4, Eq.(7)], it is worth noticing that (11) is equivalent to [4, Eq.(7)] but it is relatively simple since it does not involve the exponential integral $\mathrm{Ei}(\cdot)$ function [11, Eq. (8.211)]. In addition, from a numerical convergence stand-point [14], (11) is numerically and computationally more efficient than [3, Eq.(6)]. For instance, Chebyshev-Gauss quadrature requires less number of terms when applied onto (11), specifically with comparison to [3, Eq.(6)] and [4, Eq.(7)].

---

[2] Note that the MGF and its derivative are available in closed form for fading distributions commonly used in the literature (see for example [1, Table II-V] and [2, Table I-III]).

$$\mathcal{Z}_3(s) = -\exp(-s)\left(\log^3(s) + 3\mathrm{E}\log^2(s) - \left(\tfrac{\pi^2}{2} - 3\,\mathrm{E}^2\right)\log(s) - \left(\tfrac{\pi^2}{2} - \mathrm{E}^2\right)\mathrm{E} + 2\,\zeta(3)\right). \quad (16)$$

$$\mathcal{Z}_4(s) = \exp(-s)\left(\log^4(s) + 4\mathrm{E}\log^3(s) - (\pi^2 - 6\mathrm{E}^2)\log^2(s) + \left(4\mathrm{E}^3 - 2\pi^2\mathrm{E} + 8\,\zeta(3)\right)\log(s) + \mathrm{E}^4 - \pi^2\mathrm{E}^2 + 8\mathrm{E}\,\zeta(3) + \tfrac{\pi^4}{60}\right). \quad (17)$$

In wireless communication theory and statistics, the variance of the channel capacity (i.e., $\mathbb{V}[\log(1+\gamma_{end})]$ where $\mathbb{V}[\cdot]$ denotes the variance operator) is a measure of how far the channel capacity lies from the ergodic capacity. In particular, the variance is related to the first and second-order statistics of the channel capacity by $\mathbb{V}[\log(1+\gamma_{end})] = \mu_2 - \mu_1^2$. Its normalization with respect to the ergodic capacity, i.e.,

$$\mathbb{D}[\log(1+\gamma_{end})] = \mu_2/\mu_1 - \mu_1 \quad (12)$$

also further explores the amount of dispersion (AoD) in the channel capacity, and the AoD takes positive values smaller than one. More specifically, when the channel quality (i.e., diversity) increases, $\mathbb{D}[\log(1+\gamma_{end})]$ approaches to zero while it approaches to one otherwise. Accordingly, we can define the reliability percentage of the signal throughput as $\mathbb{R}[\log(1+\gamma_{end})] = 100\,(1 - \mathbb{D}[\log(1+\gamma_{end})])$, i.e.,

$$\mathbb{R}[\log(1+\gamma_{end})] = 100\,(\mu_1 - \mu_2/\mu_1 + 1). \quad (13)$$

Accordingly and demonstratively, substituting $n=2$ into (3) and performing some simple algebraic manipulations using [13, Eq. (07.34.06.0017.01)] results in

$$\mathcal{Z}_2(s) = \exp(-s)\left(\log^2(s) + 2\mathrm{E}\log(s) + \mathrm{E}^2 - \tfrac{\pi^2}{6}\right). \quad (14)$$

Substituting (14) into (2), the second-order statistics of the channel capacity can be in general written as

$$\mu_2 = \int_0^\infty \exp(-s)\left(\log^2(s) + 2\mathrm{E}\log(s) + \mathrm{E}^2 - \tfrac{\pi^2}{6}\right) \times \left\{\mathcal{M}_{\gamma_{end}}(s) - \frac{\partial}{\partial s}\mathcal{M}_{\gamma_{end}}(s)\right\} ds. \quad (15)$$

Finally, the variance and the AoD in the channel capacity can be obtained by means of (11) and (15).

Additionally, the other two important statistical metrics are the skewness and kurtosis. More precisely, the skewness, which is given by $\mathbb{S}[\log(1+\gamma_{end})] = (\mu_3 - \mu_1^3)/(\mu_2 - \mu_1^2)^{3/2}$, is a measure of the degree of asymmetry for the distribution of the channel capacity, while the kurtosis, which is given by $\mathbb{K}[\log(1+\gamma_{end})] = (\mu_4 - \mu_1^4)/(\mu_2 - \mu_1^2)^2$, is the degree of peakedness of the channel capacity around the ergodic capacity. When the number of branches of the MRC receiver (i.e., diversity) increases, the skewness of the channel capacity approaches to zero, and the kurtosis of the channel capacity gets closer to the kurtosis of the channel capacity in the case of that $\gamma_{end}$ follows Gaussian distribution. It is clear that, in order to obtain the skewness and kurtosis, the third- and fourth-order statistics can be readily obtained by substituting the simplified auxiliary functions given at the top of this page, in which $\zeta(\cdot)$ denotes the zeta function [11, Eq.(9.513/1)].

### III. APPLICATION IN CORRELATED ENVIRONMENT

Let $\boldsymbol{r}$ represent a $L \times 1$ complex Gaussian random vector, that denotes the complex channel gains, with mean $\boldsymbol{\eta} = [\eta_1, \eta_2, \ldots, \eta_L]^T$ and non-singular covariance matrix $\boldsymbol{C} = \mathbb{E}\left[(\boldsymbol{r}-\boldsymbol{\eta})(\boldsymbol{r}-\boldsymbol{\eta})^H\right]$, where the superscript $T$ and $H$ denote the transposition and Hermitian transposition, respectively. Herein, the mean vector represents the line-of-sight conditions in the fading environment. Further, the instantaneous SNR $\gamma_{end}$ at the output of the $L$-branch MRC receiver whose diversity branches are correlated regarding the covariance matrix $\boldsymbol{C}$ is given by $\gamma_{end} = \boldsymbol{r}^H\boldsymbol{r}$. With the situation that the fading figure (i.e., diversity order) is common to all the diversity branches, the MGF $\mathcal{M}_{\gamma_{end}}(s) = \mathbb{E}[\exp(-s\,\gamma_{end})] = \mathbb{E}[\exp(-s\,\boldsymbol{r}^H\boldsymbol{r})]$ can be in general, as the other contribution of this letter, given by

$$\mathcal{M}_{\gamma_{end}}(s) = \frac{\exp\!\left(-s\,\Omega\,\boldsymbol{\lambda}^H\left[\boldsymbol{I} + s\tfrac{\Omega}{m}\boldsymbol{R}\right]^{-1}\boldsymbol{\lambda}\right)}{\left|\boldsymbol{I} + s\tfrac{\Omega}{m}\boldsymbol{R}\right|^m} \quad (18)$$

which we termed as the MGF of the sum of *the correlated non-central chi-squared distributions / the correlated generalized Rician distributions*. In (18), $\boldsymbol{R}$ is the $L \times L$ normalized covariance matrix defined by $\boldsymbol{R} = \boldsymbol{C}/(\boldsymbol{\eta}^H\boldsymbol{\eta} + \mathrm{Tr}(\boldsymbol{C}))$ where $\mathrm{Tr}(\cdot)$ denotes the trace operator, and $\boldsymbol{\lambda}$ is the $L \times 1$ normalized mean vector defined by $\boldsymbol{\lambda} = \boldsymbol{\eta}/\sqrt{\boldsymbol{\eta}^H\boldsymbol{\eta} + \mathrm{Tr}(\boldsymbol{C})}$. The parameter $\Omega$ corresponds to the average SNR at the output of the MRC receiver[3] and the parameter $m$ denotes the diversity order of the signal recovered by each branch. Furthermore, the notations $\boldsymbol{I}$, $|\cdot|$ and $[\cdot]^{-1}$ in (18) denote the $L \times L$ identity matrix, matrix determinant and matrix inversion, respectively. At the point regarding the versatility of (18), it is useful mentioning that the MGF given by (18) simplifies to the MGF of the MRC receiver over correlated Nakagami-$m$ fading channels [15, Eq. (9.219)] when the mean vector $\boldsymbol{\eta} = 0$. Indeed, for $m=1$ and $\boldsymbol{\eta} = 0$, it reduces to the MGF over Rayleigh fading channels. Furthermore, substituting $m=1$ into (18) and then performing some simple algebraic manipulations results in the MGF of the MRC receiver over correlated Rician fading channels [3, Eq. (5)]. In addition, when the diversity order $m$ of the branches gets much larger (i.e., $m \to \infty$), (18) simplifies to $\mathcal{M}_{\gamma_{end}}(s) = e^{-s\,\Omega\,\boldsymbol{\lambda}^H\boldsymbol{\lambda}}$ which is the MGF of the MRC receiver operating in additive white Gaussian noise (AWGN) channels.

In order to obtain the higher-order capacity of the channel capacity, the first derivative of the MGF $\mathcal{M}_{\gamma_{end}}(s)$ given in (18) is required as per Theorem 1. Performing some algebraic manipulations and using [11, Eq.(13.31/5)], we obtained

$$\frac{\partial}{\partial s}\mathcal{M}_{\gamma_{end}}(s) = -\Omega\,\frac{\exp\!\left(-s\,\Omega\,\boldsymbol{\lambda}^H\left[\boldsymbol{I}+s\tfrac{\Omega}{m}\boldsymbol{R}\right]^{-1}\boldsymbol{\lambda}\right)}{\left|\boldsymbol{I}+s\tfrac{\Omega}{m}\boldsymbol{R}\right|^m} \times \left\{\mathrm{Tr}\!\left([\boldsymbol{I}+s\tfrac{Ω}{m}\boldsymbol{R}]^{-1}\boldsymbol{R}\right) + \boldsymbol{\lambda}^H\left[\boldsymbol{I}+s\tfrac{\Omega}{m}\boldsymbol{R}\right]^{-2}\boldsymbol{\lambda}\right\}. \quad (19)$$

---

[3] Note that the average SNR $\Omega$ at the output of the MRC receiver is $\Omega = \sum_{\ell=1}^{L} \bar{\gamma}_\ell$ where $\bar{\gamma}_\ell$ denotes the average SNR of the $\ell$th diversity branch. The average SNRs of the diversity branches, i.e., $\bar{\boldsymbol{\gamma}} = [\bar{\gamma}_1, \bar{\gamma}_2, \ldots, \bar{\gamma}_L]^T$ may not be balanced due to both the covariance matrix $\boldsymbol{C}$ and the mean vector mean $\boldsymbol{\eta}$, and specifically are given by $\bar{\boldsymbol{\gamma}} = \Omega\,\mathrm{diag}\left(\boldsymbol{R} + \boldsymbol{\lambda}\boldsymbol{\lambda}^H\right)$ where $\mathrm{diag}(\cdot)$ denotes the diagonal of the matrix argument.

$$\mu_n = \int_0^\infty \mathcal{Z}_n(s)\exp\bigl(-s\,\Omega\,\boldsymbol{\lambda}^H\bigl[\boldsymbol{I}+s\tfrac{\Omega}{m}\boldsymbol{R}\bigr]^{-1}\boldsymbol{\lambda}\bigr)\,\Bigl|\boldsymbol{I}+s\tfrac{\Omega}{m}\boldsymbol{R}\Bigr|^{-m}\Bigl\{1+\operatorname{Tr}\bigl(\bigl[\boldsymbol{I}+s\tfrac{\Omega}{m}\boldsymbol{R}\bigr]^{-1}\boldsymbol{R}\bigr)+\boldsymbol{\lambda}^H\bigl[\boldsymbol{I}+s\tfrac{\Omega}{m}\boldsymbol{R}\bigr]^{-2}\boldsymbol{\lambda}\Bigr\}\,ds. \quad (20)$$

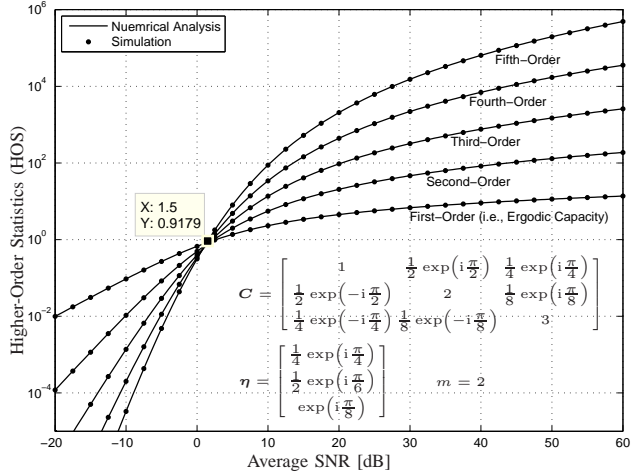

Fig. 1. Higher-order statistics of the channel capacity for a 3-branch MRC receiver in correlated generalized fading environments.

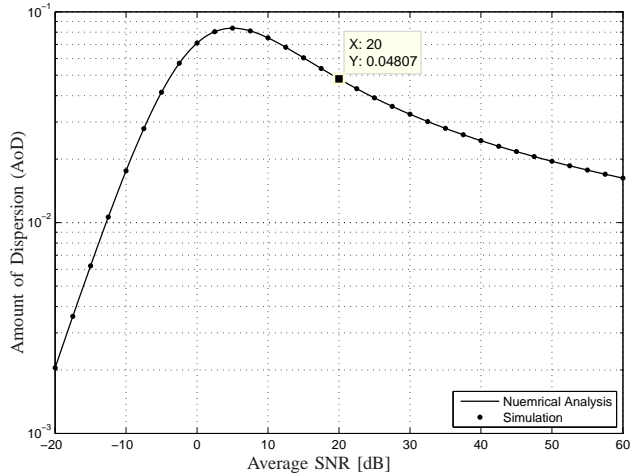

Fig. 2. Amount of dispersion of the channel capacity for the 3-branch MRC receiver considered in Fig. 1.

Finally, substituting (18) and (19) into (2), the HOS of the channel capacity of the $L$-branch MRC receiver over correlated generalized fading channels can be readily obtained as shown in (20) at the top of this page.

In order to check analytical simplicity and accuracy, the considered covariance $\boldsymbol{C}$, mean $\boldsymbol{\eta}$ and fading figure $m$ are explicitly given in Fig. 1 for a 3-branch MRC receiver. As seen in Fig. 1, the HOS are getting closer to each other almost around $1.5\,\text{dB}$. This value determines the boundary SNR $< 1.5\,\text{dB}$ of that the low-SNR regime starts. As seen in Fig. 1, the HOS $\mu_n$ in the low-SNR regime can be characterized with $\mu_n \approx \mu_1^n$. In addition, in Fig. 2, the AoD in the channel capacity of the considered 3-branch MRC receiver is depicted. As seen in Fig. 2, the AoD distinctly increases, reaches its highest value around $5\,\text{dB}$, and then decreases as the average SNR increases. For the SNR values either much lower or much higher than the SNR the AoD peaks, the channel capacity does not fluctuate drastically, specifically conveying that the transmission throughput is reliable (i.e., even if the throughput gets either worse for low SNR values or better for high SNR values). As a consequence, for a better communication configuration, the SNR should be chosen greater than the SNR the AoD peaks. For instance, for $20\,\text{dB}$, the AoD is $0.04807$ and the reliability percentage is $95.193$, which means that the average SNR must be chosen equal to or greater than $20\,\text{dB}$ in order to reach at least $95.193\%$ reliable throughput.

## IV. Conclusion

In this letter, we proposed a novel MGF-based unified analysis of HOS of the channel capacity for the MRC receivers operating in generalized correlated fading channels, and also offered a closed-form solution for the MGF of the sum of the correlated non-central chi-squared fading distributions. To validate our exact analytical expressions, Monte Carlo simulations have been carried out, and the numerical and simulation results were shown to be in perfect agreement.